\documentclass[sigconf]{acmart}

\usepackage[most]{tcolorbox}
\usepackage{adjustbox}
\usepackage{color}
\usepackage{xcolor}
\usepackage{listings}
\usepackage{graphicx}
\usepackage[labelformat=simple]{subcaption}
\usepackage{xspace}
\usepackage{multirow}
\usepackage[ruled,vlined]{algorithm2e}
\usepackage{ulem}
\usepackage{textcomp}
\usepackage{seqsplit}
\usepackage{tabularx}
\usepackage{enumitem}

\renewcommand\footnotetextcopyrightpermission[1]{} 
\AtBeginDocument{%
  }

\setcopyright{acmlicensed}
\copyrightyear{2018}
\acmYear{2018}
\acmDOI{XXXXXXX.XXXXXXX}

\acmConference[Conference acronym 'XX]{Make sure to enter the correct
  conference title from your rights confirmation emai}{June 03--05,
  2018}{Woodstock, NY}
\acmISBN{978-1-4503-XXXX-X/18/06}

\begin{document}

\title{Unveiling Ethereum's P2P Network: The Role of Chain and Client Diversity}

\author{Jiahao Luo}
\email{luoshitou9@gmail.com}
\orcid{0009-0001-4174-9996}

\renewcommand{\shortauthors}{Trovato et al.}

\begin{abstract}
The Ethereum network, built on the devp2p protocol stack, was designed to function as a “world computer” by supporting decentralized applications through a shared P2P infrastructure. However, the proliferation of blockchain forks has increased network diversity, complicating node discovery and reducing efficiency. Ethereum mainnet nodes cannot easily distinguish between peers from different blockchains until after establishing an expensive TCP connection, encryption, and protocol handshake. This inefficiency is further worsened by client diversity, where differences in software implementations cause protocol incompatibilities and connection failures. This paper introduces a monitoring tool that tracks devp2p message exchanges and client statuses to analyze connection dynamics and protocol variations. Our findings highlight issues such as inefficiencies in node discovery and client incompatibility, including timeouts in Geth during the discovery process. The study emphasizes the need to consider chain and client diversity when assessing the health and performance of the post-merge Ethereum network.
\end{abstract}




\maketitle

\section{Introduction}

The evolution of the Web into its next phase, Web3, has ushered in a new paradigm of decentralized peer-to-peer systems, where blockchain technology plays a pivotal role in reimagining how applications and services are built and operated. In this decentralized web, Ethereum~\cite{ethereum_yellowpaper} is at the forefront, providing the foundation for secure and trustless interactions without central intermediaries. Ethereum’s architecture, particularly its reliance on the devp2p protocol stack, underpins its ability to serve as a 'world computer', allowing decentralized applications (dApps) to function seamlessly on a global scale.

This protocol stack is composed of three key layers: the node discovery protocol, the RLPx transport protocol, and various sub-protocols that are critical for node discovery, peer connections, and higher-level application interactions~\cite{ethereum_devp2p}. The node discovery and RLPx transport protocols are essential for ensuring continuous network connectivity and secure communication among peers. Meanwhile, sub-protocols, such as the Ethereum wire protocol, facilitate the decentralized application (dApp) ecosystem.

Measurement of Ethereum network behavior offers valuable insight into the robustness of client implementations and helps to assess network performance and potential security risks. Previous research has primarily focused on application-level observations, including transaction and block propagation~\cite{heoBlockExplorersPublic2021,heoPartitioningEthereumEclipsing2023}, active network topology~\cite{liTopoShotUncoveringEthereum2021,zhaoDEthnaAccurateEthereum2024}, denial-of-service (DoS) attacks in the transaction pool~\cite{liDETERDenialEthereum2021,cryptoeprint:2023/956,wang2023understanding}, and the extraction of maximum extractable value (MEV)~\cite{weintraubFlashBotPan2022,heimbachEthereumProposerBuilderSeparation2023,torres2021frontrunner}.

However, these studies largely overlook low-level peer-to-peer (P2P) message exchanges, which are critical to understanding the internal status of Ethereum clients that implement the devp2p protocol~\cite{kifferHoodEthereumGossip2021,gao2019topology,kimMeasuringEthereumNetwork2018}. Furthermore, recent Ethereum upgrades~\cite{ethereum_paris_upgrade,ethereum_merge_roadmap} have substantially altered the network, making previous observations less relevant to its current state.

Our motivation is to examine the dynamics of the P2P network with a particular focus on \textit{chain diversity} and \textit{client diversity}, two factors that significantly impact the health of the Ethereum network.

Chain diversity can be understood in two ways. First, it refers to the numerous blockchain networks that leverage Ethereum’s devp2p protocol but operate independently from Ethereum’s mainnet. Ethereum’s original vision of creating a “world computer”\cite{ethereum_smart_contracts}, encompassing projects like Swarm\cite{swarm} and Whisper~\cite{ethereum_whisper}, has expanded beyond its initial scope, with the emergence of numerous new blockchain networks (e.g., Polygon~\cite{polygon_bor}). Currently, Chainlist~\cite{chainlist} tracks more than 1,400 chains, highlighting the significant role of chain diversity within the Ethereum ecosystem.
The second aspect of \textit{chain diversity} involves peers that have not kept up with the different versions of Ethereum's network, known as forks. Forks occur when changes or updates to the Ethereum protocol are introduced, which can result in separate versions of the network. Peers that have not updated to the latest fork become incompatible with the updated protocol. To disconnect these incompatible peers and free up valuable connection slots for updated peers with public IP addresses, Ethereum introduced a mechanism called the fork identifier~\cite{eip2124,eip6122}. However, identifying these incompatible peers using the fork identifier can only be done after an expensive process: establishing a TCP connection, encrypting data with RLPx, and completing the Ethereum wire protocol handshake. The second aspect of chain diversity involves Ethereum forks.

Client diversity refers to the variety of software implementations, or “clients”, that allow nodes to participate in the Ethereum network~\cite{client_diversity}. Each client, developed by different teams, includes distinct features, optimizations, and behaviors~\cite{eip7636}. This diversity enhances network security and resilience: if one client has a vulnerability, the rest of the network can operate via other clients.

However, client diversity presents challenges. Differences in how clients interpret the Ethereum protocol can result in communication mismatches or varying transaction validation processes. These discrepancies can cause nodes to fail in establishing connections or lead to inconsistent transaction processing across network.

In this paper, we develop a modified Ethereum client and a workflow to analyze the network’s behavior. We focus on two primary aspects: (1) how network connections are affected by different chains that share Ethereum’s devp2p protocol (chain diversity), and (2) how varying client implementations (client diversity) lead to compatibility issues and differing protocol behaviors. Using a real-time analyzer, we capture and examine messages exchanged between peers and monitor the internal state of our local client. This setup allows us to study peer interactions and evaluate the stability of the network from the perspective of our local node.

\noindent \textbf{Contributions.} Our key contributions are as follows:

\begin{itemize}[leftmargin=*,topsep=1pt] 

\item  We develop an open-source tool\footnote{\url{https://github.com/learnerLj/ethereum-measurement}} that monitors inbound and outbound messages, tracking their effects on the internal state of local clients. This tool enables us to identify message errors and client incompatibilities. To the best of our knowledge, this is the first study to systematically analyze Ethereum’s P2P messages alongside client compatibility after the network’s transition from proof-of-work to proof-of-stake, known as the merge.

\item We identify node discovery and connection management issues caused by client diversity. Variations in how clients handle disconnections can mislead nodes, while differing node discovery approaches often result in enforced timeouts, particularly in Geth, Ethereum’s most widely used client.

\item We explore Ethereum’s chain diversity and find that only 78.05\% of peers use the latest Ethereum mainnet protocol, with just 24.5\% having the correct mainnet configuration. Furthermore, 12.6\% of peers are behind on critical updates, such as the most recent hard fork. Our analysis reveals a substantial shift in connection dynamics, with the proportion of mainnet peers decreasing from 54.5\% in 2018 to just 13.07\% today.

\end{itemize}

\section{Background}
The Ethereum peer-to-peer (P2P) network enables nodes to interact directly without relying on central intermediaries. To establish secure connections for decentralized applications, Ethereum uses the devp2p protocol, which offers three main functionalities: 1) a node identity scheme for identifying peers, 2) a peer discovery mechanism to locate peers’ endpoints (e.g., IP addresses, TCP, and UDP ports), and 3) public key distribution to secure communications. Ethereum enhances network robustness and fault tolerance through its bespoke devp2p protocol stack, which integrates Kademlia, a distributed hash table (DHT) based on UDP~\cite{maymounkov2002kademlia}.

\subsection{Node Discovery}
The Ethereum network currently relies on version 4 of the discovery protocol (discv4) for node discovery. Each node generates a 256-bit private key and a corresponding 512-bit Elliptic Curve Digital Signature Algorithm (ECDSA) public key~\cite{johnson2001elliptic}, truncated by removing the \texttt{0x04} prefix. The node’s public key is then hashed using Keccak256~\cite{apriani2021performance} to produce a unique identifier, known as a node ID, derived from the first 80 bits of the hash.

A node’s essential record includes its node ID, IP address, TCP port, and UDP port. To measure the logical proximity of two nodes, the protocol computes the bitwise XOR (exclusive OR) of their node IDs: $distance(node_1, node_2) = node_1\_{id} \oplus node_2\_{id}$. This metric helps organize nodes in the DHT.

Upon receiving a \texttt{FindNode} request targeting a specific public key, a node responds with a \texttt{Neighbors} message that lists the 16 closest nodes from its table~\cite{discv4_findnode}. Once a node’s endpoint is discovered, an Ethereum Node Record (ENR) request is issued to obtain the node’s compressed public key, which is required to establish a secure TCP connection. Node discovery uses UDP to minimize overhead and provide proof of endpoint availability. Nodes respond only to requests from peers that have recently sent a ping (within the last 12 hours), ensuring that only bonded nodes (nodes that have exchanged pings) receive \texttt{Neighbors} messages (typically 1,280 bytes) or \texttt{ENR responses} (approximately 120 bytes). This helps mitigate potential traffic amplification attacks.

\subsection{Distributed Hash Table} 
Each node keeps a table of other nodes, storing details like their network address (endpoint) and node ID. These records are grouped into "buckets" based on how similar (or close) their node IDs are to the local node's ID. Since it's very rare for nodes to have highly similar IDs (the odds are less than $1/2^{16}$ of having more than 16 identical prefix bits), Geth simplifies this by using just 17 buckets for storing nodes that are closer in ID~\cite{go_ethereum}. Each bucket holds up to 16 records, and they are sorted by how recently the node responded.

Each node maintains a table of other nodes, categorized by their node IDs. These records are organized into “buckets” based on how similar their node IDs are to the local node’s ID. Given the low probability of nodes sharing highly similar IDs (less than $1/2^{16}$ chance of matching more than 16 prefix bits), Geth simplifies node storage by using only 17 buckets for nodes closer in ID~\cite{go_ethereum}. Each bucket can hold up to 16 records, sorted by the most recent response time.

To ensure table accuracy, the node periodically pings the last node in a randomly chosen bucket to verify its availability (a process called endpoint proof). If the node responds with a pong, it confirms the node’s continued reachability and updates any changes to the node’s network address.

Additionally, the node refreshes its table by sending \texttt{FindNode} requests to random nodes, discovering new peers and updating its internal records~\cite{henningsen2019eclipsing}.

\subsection{RLPx Handshake}
The RLPx transport protocol is Ethereum’s mechanism for securely transmitting messages between nodes. It employs a random selection process to ensure nodes have an equal chance of connecting to others, regardless of the local peer table’s structure.

\begin{figure}[H]
    \centering
    \includegraphics[width=0.58\linewidth]{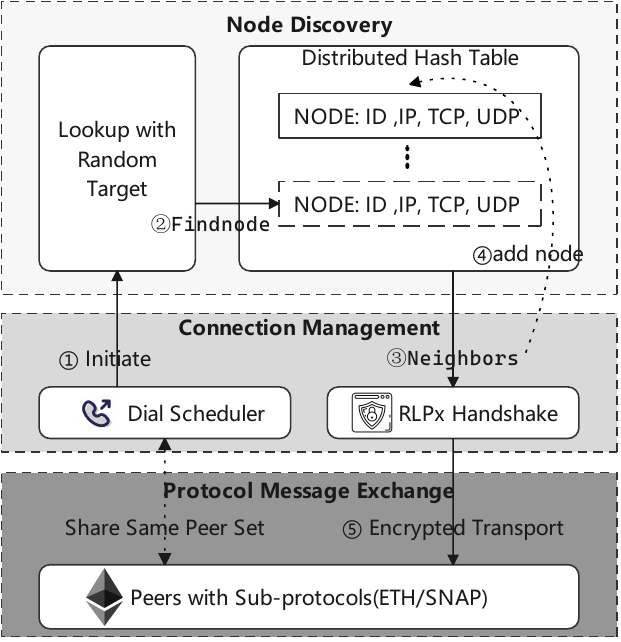}\vspace{-8pt}
    \caption{Overview of the Ethereum network architecture.}
    \label{fig:network-arch}
\end{figure}

As illustrated in Figure~\ref{fig:network-arch}, nodes are not selected directly from local table but instead from the peer tables of other known nodes~\cite{henningsen2019eclipsing,go_ethereum_lookup_2023}. This means that while a node may maintain a list of peers, it often connects to the peers of its peers, promoting network randomness and preventing inference about node relationships. Even though peers in the table are labeled “neighbors,” application-level communication between them is not guaranteed.

The connection process consists of two stages: an encryption handshake to secure communication, followed by a protocol handshake where nodes exchange information about supported protocols (e.g., eth/68~\cite{ethereum_eth_protocol_2024}, snap/1~\cite{ethereum_snap_protocol_2024}). Only nodes that support the same protocols proceed to further communication. Once the secure TCP connection is established, messages are managed via the RLPx handler for simple exchanges (like ping and pong) or through sub-protocols for more complex tasks such as snapshot synchronization and transaction broadcasting.

Figure~\ref{fig:network-arch} illustrates how nodes are discovered and integrated into the network.

\vspace{-8pt}
The connection process starts when the Dial Scheduler checks for available slots and selects a node from the discovery table (\textcircled{1}). A lookup task is then initiated by sending a \texttt{FindNode} message to nodes closest to a randomly chosen target (\textcircled{2}). Upon receiving the request, these nodes respond with a \texttt{Neighbors} message listing up to 16 nodes and their endpoints (\textcircled{3}). These nodes are then added to the local Distributed Hash Table (DHT) (\textcircled{4}) and considered for connection via an RLPx handshake (\textcircled{5}). If a remote node supports compatible sub-protocols and chain parameters, it is added to the peer set, avoiding unnecessary retries by the Dial Scheduler.

\section{Methodology and Implementation}

\subsection{Design}
This study adopts a comprehensive approach to examine protocol interactions and performance issues within the Ethereum network. By monitoring both packet exchanges and internal client states (e.g., failures, retries), we systematically capture Ethereum’s network behavior. Our focus is on key components of the protocol stack—such as peer discovery, endpoint verification, and sub-protocol exchanges—to extract relevant data for analysis.

\begin{figure*}[htbp]
    \centering
    \includegraphics[width=0.7\linewidth]{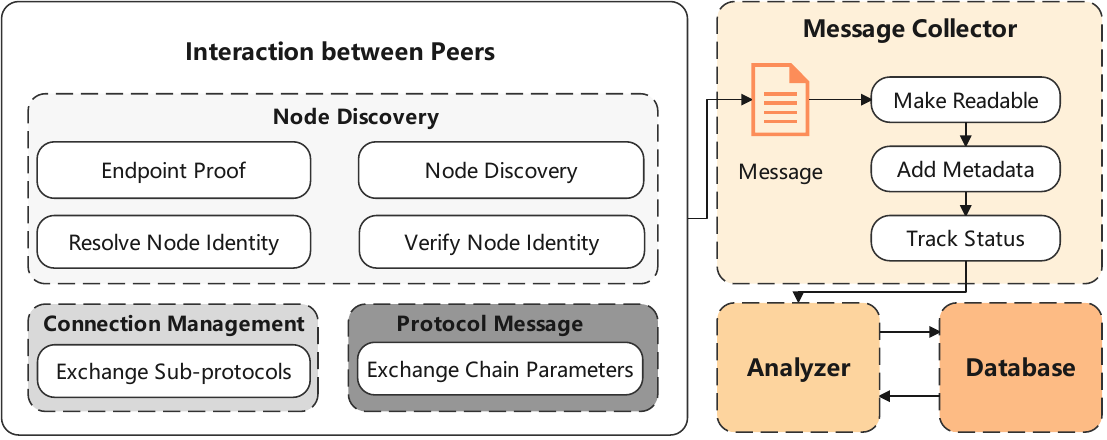}\vspace{-8pt}
    \caption{The architecture of measurement system design.}
    \label{fig:measure-arch}
\end{figure*}

The architecture of our measurement system, depicted in Figure~\ref{fig:measure-arch}, comprises four main components: interaction monitoring, message collection, data analysis, and storage.

\subsubsection{Peer Interactions} 
As detailed in Appendix D, Ethereum’s peer-to-peer communication process is fundamental to maintaining network health and connectivity. Our methodology involves instrumenting the Ethereum client code based on protocol specifications to intercept and log relevant data during peer interactions. This instrumentation allows us to track key points in the peer discovery and message exchange processes, enabling packet-level monitoring and system-level logging. Additionally, we log client state transitions, failures, and retries, offering a holistic view of interaction dynamics across Ethereum clients. This data provides insight into both the protocol’s robustness and potential inefficiencies in a diverse network.

\subsubsection{Message Collector} 
The message collector component is designed to capture and decode all messages exchanged during peer interactions. Each message is decoded from its byte-encoded form and enriched with metadata (e.g., message ID, timestamp) to maintain context. The collector also links related interactions, such as a \texttt{Neighbors} response being connected to the original \texttt{FindNode} request, enabling precise tracking of peer exchanges and compliance with protocol specifications.

\subsubsection{Analyzer and Database} 
The analyzer processes the collected data by applying user-defined rules, extracting meaningful insights based on the study’s objectives. Beyond analyzing message exchanges, the analyzer integrates internal node data (e.g., available peer slots, pending connections), allowing for real-time network performance monitoring. The system stores historical data in a dedicated database for long-term analysis and further refinement of rules, improving the efficiency of future measurements.

\subsection{Implementation}
This section provides details on the system setup for our measurement infrastructure, along with practical and reproducible configurations and implementations.

\subsubsection{Analyzer}
The core function of our system is to collect and analyze the status of Ethereum peers. This begins with gathering key metadata about each peer, including its node ID, IP address, and TCP/UDP ports, as well as identifying the sub-protocols each peer supports. The steps below outline how the analyzer processes and categorizes this data to assess network conditions.

\textit{\textbf{Step 1: Peer Identity and Endpoint Discovery.}}
Using the node discovery protocol, we obtain a peer’s identity, which includes its public key and endpoint. Peers can be discovered through three methods:

\begin{itemize}[leftmargin=*,topsep=1pt]
    \item \textbf{Receiving a \texttt{Ping}:} A \texttt{Ping} message verifies that the node is actively participating in the discovery protocol. The message contains the node’s endpoint and node ID (derived from its signature), confirming the node is “bonded” and ready for further interaction.

    \item \textbf{Receiving a \texttt{Neighbors} Message:} This message includes multiple node identifiers and endpoints, which the message collector decodes and forwards for analysis.

    \item \textbf{Receiving an RLPx \texttt{Hello}:} When a node initiates a connection, the \texttt{Hello} message provides its identity directly from the socket, including details like node ID and supported sub-protocols.

\end{itemize}

\textit{\textbf{Step 2: Sub-protocol Discovery.}}
The \texttt{Hello} message also contains essential information about the sub-protocols that a peer supports. For peers that we actively dial, resolving the endpoint beforehand means that metadata (e.g., IP and node ID) need not be re-logged if already captured. The dial scheduler updates records with sub-protocol information for both inbound and outbound connections. Appendix E contains detailed examples of peer information processing.

\textit{\textbf{Step 3: Sub-protocol Parameter Validation.}}
Even when two peers support the same sub-protocols, they may operate on incompatible parameters (e.g., version, network ID). Using Ethereum wire protocol version 68, we compare these parameters by analyzing the \texttt{Status} message received from peers. This helps ensure compatibility, updating records with chain parameters such as network ID and forkID.

Future iterations of the analyzer may include custom modules for monitoring specific statuses, such as identifying peers behind NATs or firewalls based on \texttt{Ping} response patterns. This would provide deeper insights into the network’s topology.
\subsubsection{Client and System Configuration}
To optimize the measurement system, we made several adjustments to client and system settings, improving the efficiency of node discovery, connection establishment, bandwidth usage, and data collection. Key configuration changes include: 1) Modifications to the discovery protocol to enhance peer identification; 2) Adjustments to TCP buffer sizes to improve message handling; 3) Logging optimizations to capture relevant data with minimal overhead; 4) Setting up a MongoDB instance for high-performance storage and retrieval of large datasets.
Full details of these configurations are outlined in Appendices A and B, where we provide a step-by-step guide to replicating our setup for other research or operational purposes.

\vspace{-4pt}
\section{Measurement Results}
\subsection{Message Collection}

Over the course of the study, we collected more than 8 billion messages, with \texttt{FindNode} and its prerequisite \texttt{Ping} messages constituting approximately 91\% of the total. This indicates that our node discovery process was highly efficient, as every \texttt{FindNode} request requires a preceding \texttt{Ping} to verify the endpoint. Table~\ref{tab:msg-count} provides a summary of the different types of messages captured during the study. The large number of \texttt{FindNode} messages demonstrates that our nodes actively pushed the network discovery process to its limits, consistently identifying peers that responded with verified endpoints.

\vspace{-8pt}
\begin{table}[H]
    \centering
    \caption{Number of collected messages, in millions}\vspace{-8pt}
    \begin{adjustbox}{width=\columnwidth}
        \begin{tabular}{@{}lccccccc@{}}
            \toprule
            \multirow{2}{*}{\textbf{Type}}  & \multicolumn{4}{c}{\textbf{Discovery}} & \multicolumn{2}{c}{\textbf{RLPx Messages}} & \multicolumn{1}{c}{\textbf{Eth}} \\
            \cmidrule(r){2-5} \cmidrule(l){6-7}
             & \textbf{Ping} & \textbf{Pong} & \textbf{FindN.} & \textbf{Neigh.}& \textbf{Disc} & \textbf{Hello} & \textbf{Status} \\
            \midrule
            \textbf{Count} & 3,735 & 39  & 3,996  & 172 & 78 & 366 & 127 \\
            \bottomrule
        \end{tabular}
    \end{adjustbox}
    \label{tab:msg-count}
\end{table}
\vspace{-8pt}

\begin{figure}[H]
    \centering
    \includegraphics[width=0.8\linewidth]{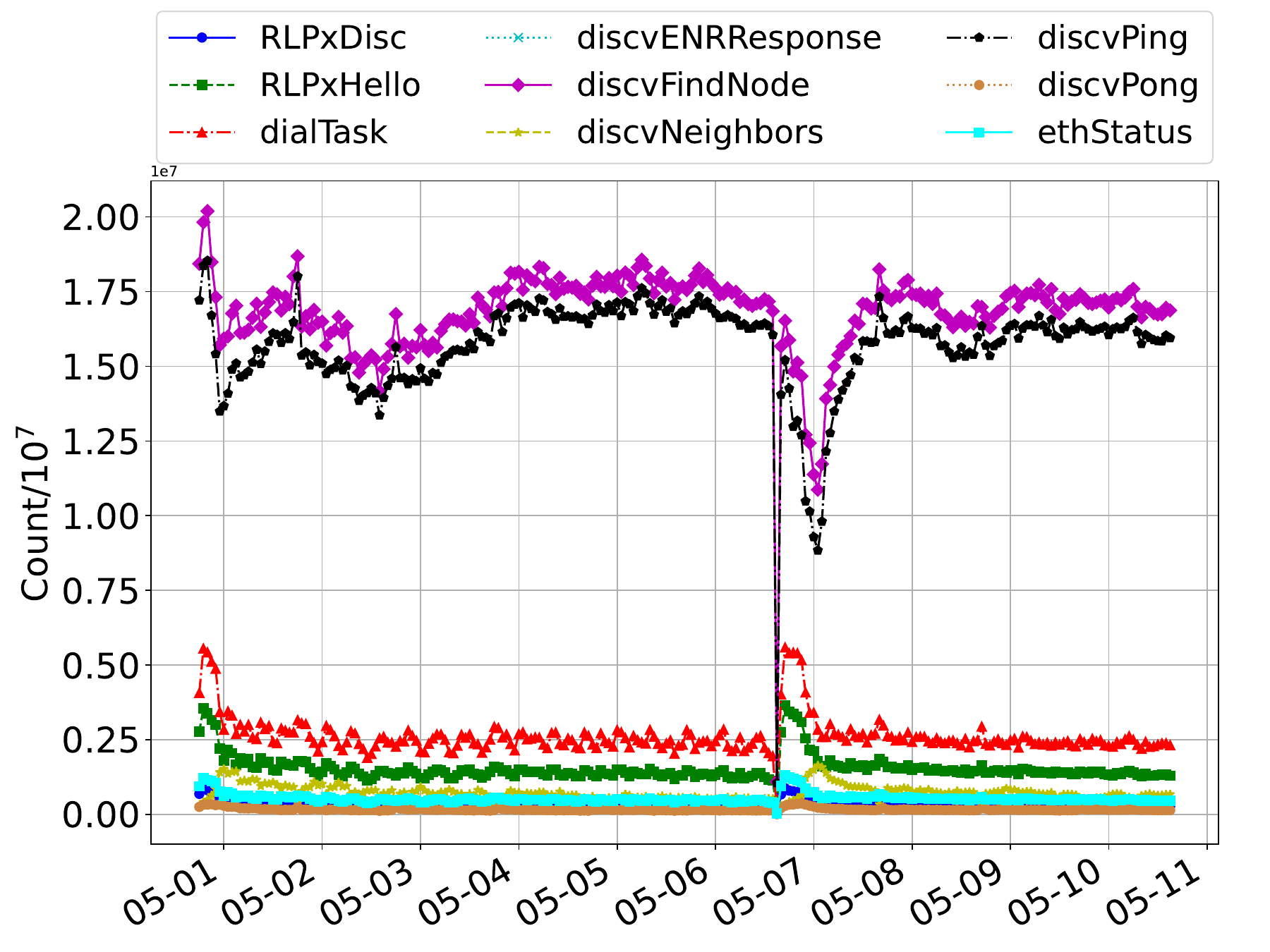} \vspace{-8pt}
    \caption{Messages and tasks per hour, including a brief shutdown on May 7th.}
    \label{fig:msg-cnt}
\end{figure}
\vspace{-8pt}

Our system initiated approximately 2.64 million connection attempts per hour and about 16.9 million node discovery attempts per hour. Compared to previous studies, such as Kim et al.\cite{kimMeasuringEthereumNetwork2018}, our discovery rate was significantly higher, indicating improved reachability across the Ethereum network. To avoid overwhelming the network or causing amplification, connection attempts were randomized and restricted to no more than 10 attempts per peer per minute. Figure\ref{fig:msg-cnt} illustrates the hourly rates of messages and tasks, along with a brief system shutdown on May 7th.

We identified 743,739 unique peers, each recorded with basic metadata such as node ID, IP address, and TCP/UDP ports. These peers were categorized based on their interaction with our node, with three primary classes: 1) Class A: Nodes that provided only basic endpoint data and did not engage in sub-protocol exchanges; 2) Class B: Nodes that exchanged sub-protocols but were ultimately incompatible; 3) Class C: Nodes that were fully compatible and actively engaged in sub-protocol exchanges.
Table~\ref{tab:peer-category-count} provides a summary of peer classifications, illustrating the diversity of interactions. This classification helped us evaluate the robustness and behavior of Ethereum under varying conditions, shedding light on how chain and client diversity impact peer connectivity.

The large number of Class A peers suggests that while many nodes are discoverable in the DHT of network, a significant portion do not engage beyond basic endpoint verification. The smaller number of Class C peers indicates that full compatibility between nodes remains relatively rare, likely due to the diverse range of Ethereum client implementations and varying network configurations.

\vspace{-8pt}
\begin{table}[H]
\centering
    \caption{Classification of collected peers}\vspace{-8pt}
    \begin{adjustbox}{width=\columnwidth}
        \begin{tabular}{@{}clrc@{}}
        \toprule
        \textbf{Class} & \textbf{Description} & \textbf{Count} & \textbf{Overall Total}\\
        \midrule
        A & No sub-protocol exchange & 674,945 & \multirow{3}{*}{743,739}\\
        B & Exchanged, not compatible & 17,707 & \\
        C & Compatible sub-protocols & 51,087 & \\
        \bottomrule
        \end{tabular}
    \end{adjustbox}
    
    \label{tab:peer-category-count}
\end{table}
\vspace{-8pt}

\subsection{False IP Address}
During data collection, we identified 7,934 out of 743,739 peers using private IP addresses, with 98.65\% set to 127.0.0.1 and 0.44\% to 0.0.0.0. These fallback IP addresses are typically reserved for localhost communications, making their presence in external peer discovery highly unusual. Upon reviewing our measurement framework and analyzing the message exchanges, we found that these peers were erroneously transmitting their external IP addresses, incorrectly setting the \texttt{fromIP} field of the \texttt{Ping} message to 127.0.0.1.

Despite the fallback IP address being set in the node discovery protocol’s \texttt{Ping} message, our responses were successfully directed to the peers’ external, reachable IP addresses. Further investigation pointed to a potential issue in Geth’s UPnP-NAT functionality, which defaults to using the local IP address for interactions when NAT traversal fails.

\textbf{NAT Complexity Leading to Private IP Addresses.}
To determine whether an implementation bug was responsible for this issue, we examined the interaction logs. Out of the 7,934 peers with private IP addresses, only 156 exchanged sub-protocols via the RLPx protocol. Most of these peers were using clients such as Besu (28.21\%), Geth (23.72\%), and OpenEthereum (18.59\%), indicating that the majority of peers with fallback IPs were unreachable. Those reachable primarily used Besu (v24.1.2, v24.3.3), Worldland (v1.05), Go-x1 (v1.1.5), OpenEthereum (v3.3.3), Reth (v0.1.0-alpha.21), and Nethermind (v1.25.4).
The developers of Besu acknowledged this issue and resolved it in versions v24.1.2 and v24.3.3 through PR\#6225 and PR\#6439~\cite{besuPR6225,besuPR6439}, while Reth addressed the issue earlier with multiple changes~\cite{reth_issue_4222,reth_pull_4224,reth_pull_4268,reth_pull_4849}. Despite these fixes, many clients remain unpatched and continue to use outdated versions.
Beyond fallback IP addresses, we also identified other private IP ranges, such as 10.193.x.x and 192.168.x.x. CoreGeth (v1.12.19) similarly sets the \texttt{fromIP} field to a private IP behind NAT~\cite{github-core-geth}, with forks like ETCMC~\cite{etcmc} inheriting the bug.

These findings were reported to the developers of affected clients, and our observations align with earlier vulnerabilities identified by Yi et al.~\cite{yi2022blockscope}. The persistence of these issues highlights a gap in the formal specification for the Ethereum node discovery protocol, which lacks detailed guidance on handling NAT traversal despite requiring the \texttt{Ping} message to include the sender’s IP.

\textbf{Potential Impact on Distributed Hash Table and Peer Scoring.}
The \texttt{Ping} message in Ethereum’s node discovery protocol serves as a “heartbeat” to confirm a node’s liveness, pushing the sender to the top of the distributed hash table bucket. This mechanism ensures that the DHT remains populated with active peers.
Geth retrieves the TCP port from the \texttt{Ping} message but relies on the underlying UDP socket for correct IP address and UDP port. This ensures that DHT management and peer interactions are based on reliable data. However, the \texttt{fromIP} field in discv4 \texttt{Ping} messages is not typically used for routing, raising questions about whether all client developers account for this discrepancy. If clients rely on the unverified IP addresses from \texttt{Ping} messages, this could contaminate the DHT, leading to inefficient routing and reduced connectivity.

Inaccuracies in IP addresses could also affect peer scoring mechanisms in clients like Besu and Nethermind. Peers consistently presenting non-routable IPs may receive lower scores, potentially leading to disconnection. Since peer scoring plays a critical role in connection management, faulty \texttt{Ping} messages could increase the risk of network fragmentation. While the full extent of these issues on overall network stability remains unclear, these bugs present a potential threat to efficient network operations.

\subsection{Evolving Reasons for Disconnects}
The RLPx transport protocol not only secures communication between Ethereum nodes but also manages disconnections between peers. These disconnections can result from failed connection attempts or sub-protocol-related issues. During our most recent measurement phase, we captured 78.4 million \texttt{Disconnect} messages.

\begin{table}[H]
    \centering
    \caption{Peer disconnect statistics during measurement}\vspace{-8pt}
    \begin{adjustbox}{width=\columnwidth}
        \begin{tabular}{@{}lrrr@{}}
            \toprule
             Reason & Received & Sent & Percentage (\%) \\
            \midrule
            Disconnect requested & 12,262,368 & 0 & 22.93 \\
            TCP sub-system error & 16,786,854 & 0 & 31.39 \\
            Breach of protocol & 548,827 & 0 & 1.03 \\
            Useless peer & 677 & 22,239,246 & 41.58 \\
            Too many peers & 1,519,750 & 0 & 2.84 \\
            Already connected & 1,831 & 0 & <0.01 \\
            Client quitting & 2,328 & 0 & <0.01 \\
            Ping timeout & 17,541 & 0 & 0.03 \\
            Subprotocol reason & 105,876 & 0 & 0.20 \\
            \bottomrule
        \end{tabular}
    \end{adjustbox}
    \label{tab:disconnect_stats}
\end{table}

\textbf{Useless and Unreachable Peers Predominate.}
As shown in Table~\ref{tab:disconnect_stats}, the most common reason for disconnections was Useless peer, accounting for 41.58\% of all disconnections. This category includes peers that either had no compatible sub-protocols with our nodes or failed to respond to requests. For instance, the SNAP sub-protocol disconnects peers when required blocks are not fetched, ensuring efficient chain synchronization.

The second most common reason was TCP sub-system error at 31.39\%, which indicates socket connection failures caused by unrecoverable errors, often due to network address translations (NATs) or firewall settings. The Disconnect requested category (22.93\%) typically refers to manual disconnection commands via the admin API in Geth. However, we observed that Nethermind often categorizes block-fetch failures under this reason, diverging from Geth’s classification, where such failures would fall under Useless peer. This inconsistency suggests that relying on disconnect reasons for triaging peers may not always be reliable.

Compared to data from Kim et al.~\cite{kimMeasuringEthereumNetwork2018}, the rate of Useless peer disconnections has increased significantly, from less than 1\% in 2018 to 41.58\% in 2024. This reflects the growing complexity and diversity of blockchain sub-protocols. In contrast, Too many peers disconnections dropped from 90\% in 2018 to 2.84\% in 2024, with many reclassified under TCP sub-system error or Useless peer. This shift suggests that firewall protections and sub-protocol diversity have increased the number of unreachable peers.

\textbf{Breach of Protocol.}
Although Breach of protocol disconnections account for only 1.03\% of total disconnects, they are significant, as they indicate severe protocol violations. Geth, for example, treats invalid message formats as protocol breaches—an error category not reported in Kim’s 2018 study. Our analysis found that 92.12\% of these breaches were initiated by Nethermind clients, with Geth and Reth contributing minimally. Misalignments in sub-protocol message formats, particularly among forked Ethereum clients, can pose risks to network efficiency and overall ecosystem health.

\vspace{-18pt}
\begin{figure}[h]
    \centering
    \includegraphics[width=1\linewidth]{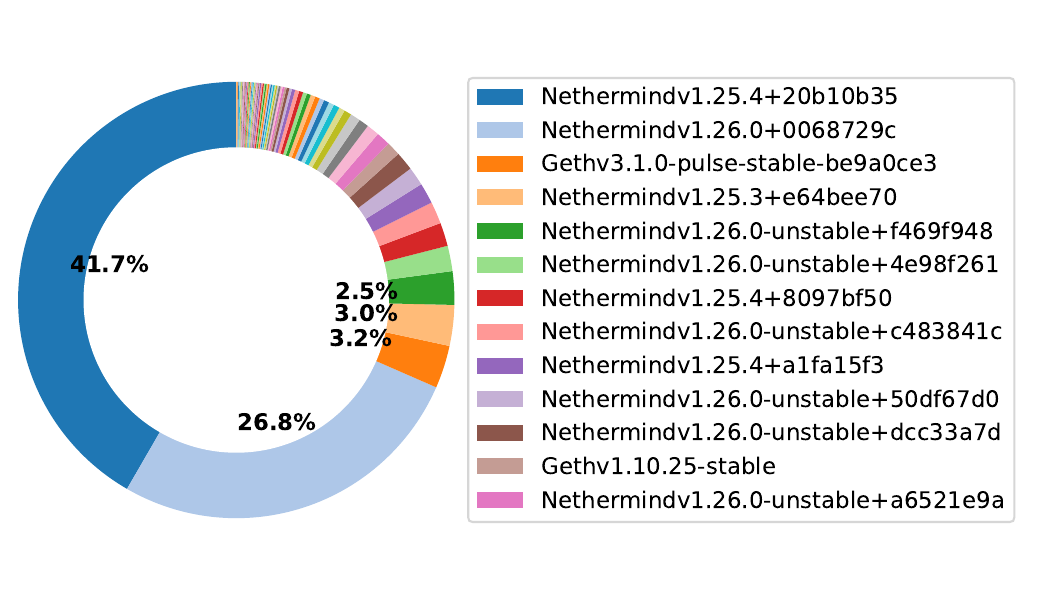}\vspace{-15pt} 
    \caption{Clients reporting breach of protocol}
    \label{fig:client-breach}
\end{figure}
\vspace{-6pt}
Upon further investigation, we found that the majority of Breach of protocol disconnects were issued by Nethermind clients, specifically versions 1.25.4 and 1.26.0. As shown in Figure~\ref{fig:client-breach}, 41.7\% of the breaches were triggered by Nethermind v1.25.4+20b10b35, followed by Nethermind v1.26.0+0068729c, which accounted for 26.8\% of the total breaches. These two versions alone represent a significant proportion of all protocol breaches, suggesting that specific versions of Nethermind are more likely to disconnect peers due to violations. Nethermind classifies several behaviors as breaches, including: (1) invalid fork ID on the mainnet, (2) invalid block gossip, and (3) failed block fetches.

Table~\ref{tab:nethermind-breach-reason} supports our findings. The first row, representing the mainnet (network ID 1), accounted for virtually none of the Breach of protocol disconnects from the 1,861 Nethermind peers. The majority occurred on alternative chains, which underscores how network-specific parameters contribute to disconnects. This divergence in disconnect reasons further highlights the inconsistency in how clients like Geth and Nethermind categorize peers. What Geth labels as a Useless peer, Nethermind flags as a Breach of protocol. Nethermind employs 45 distinct disconnect reasons, adding complexity to peer scoring systems, which could complicate uniform scoring across Ethereum clients. This issue has sparked ongoing discussions within the community~\cite{geth_29329}.

\vspace{-8pt}
\begin{table}[H]
    \centering
    \caption{Network IDs of Nethermind clients initiating breach of protocol disconnects}\vspace{-8pt}
    \begin{adjustbox}{width=\columnwidth}
    \begin{tabular}{@{}clcr@{}}
        \toprule
         Network Id & Genesis & Fork ID & Percentage (\%) \\
        \midrule
        1 & 0xd4e56740.. & 9f3d2254/0 & 0.00 \\
        17000 & 0xb5f7f912.. & 9b192ad0/0 & 46.10 \\
        100   & 0x4f1dd231.. & 1384dfc1/0 & 33.10 \\
        17000 & 0xb5f7f912.. & c61a6098/6516eac0 & 4.19 \\
        Others &           &                   & 16.61 \\
         \bottomrule
    \end{tabular}
    \end{adjustbox}
    \label{tab:nethermind-breach-reason}
\end{table}

\vspace{-8pt}
\subsection{Non-compliant Client Implementations}

A notable issue with the node discovery protocol arises from the \texttt{Neighbors} message. According to the specification, when a \texttt{FindNode} request is received, the responding peer should reply with a \texttt{Neighbors} message containing the 16 closest nodes from its local table~\cite{discv4_findnode}. However, this response is constrained by the 1280-byte size limit of a UDP datagram, which typically restricts the number of nodes to 14 when using IPv4 addresses and as few as 12 for IPv6, due to the larger address size.

\textbf{Diverse Node Counts in \texttt{Neighbors} Messages and Resulting Timeouts.}

Our measurement system analyzed over 4 billion \texttt{FindNode} messages, with more than 99.99\% being outbound and only about 5\% receiving a response. We captured approximately 172.2 million \texttt{Neighbors} messages from 74,041 unique peers, with an average of 2,285 replies per peer, demonstrating the extensive reach of our tool.

\vspace{-10pt}
\begin{figure}[H]
    \centering
    \includegraphics[width=\linewidth]{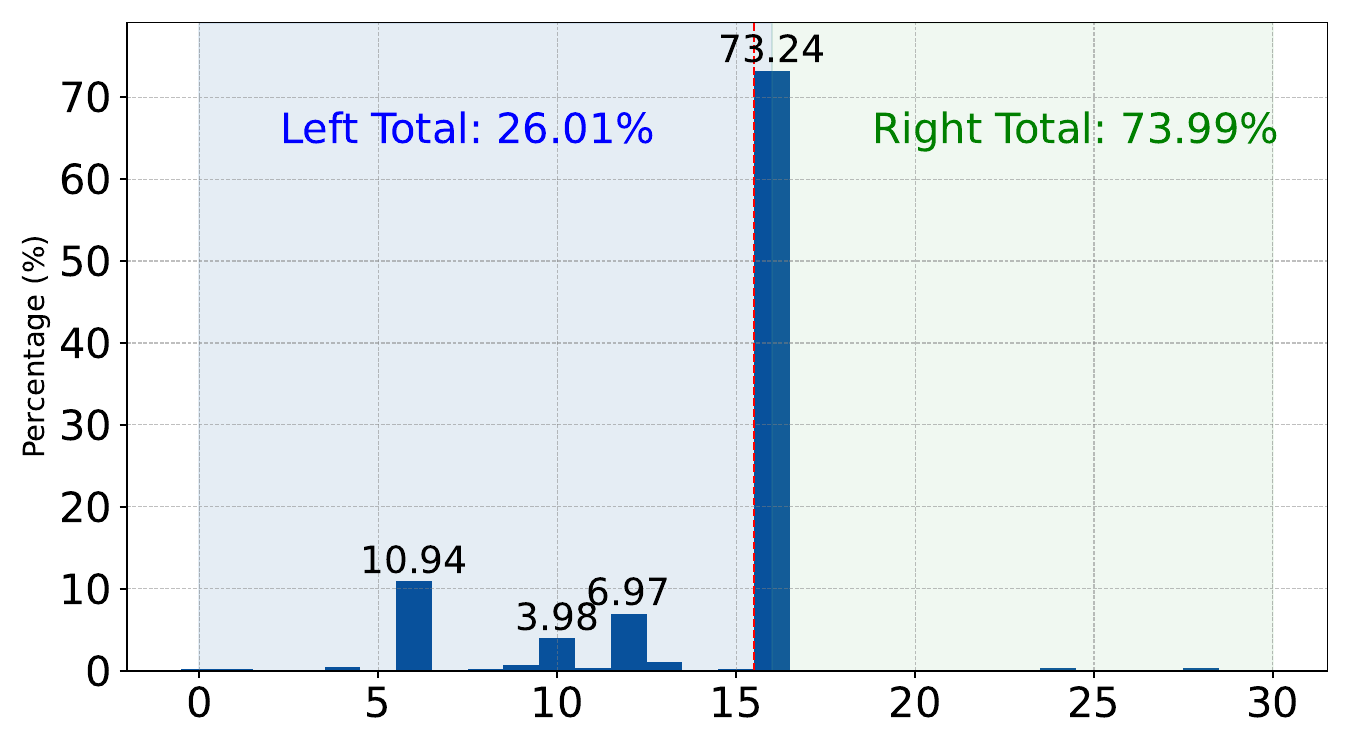}\vspace{-8pt} 
    \caption{Node lengths for a \texttt{FindNode} Request}
    \label{fig:length-neighbors}
\end{figure}
\vspace{-10pt}

An important observation is that, as shown in Figure~\ref{fig:length-neighbors}, the node lengths for a single \texttt{FindNode} request varied significantly. Roughly 73.24\% of the responses returned 16 nodes, indicating that most clients responded with the number of nodes required by the protocol specification. However, approximately 26.01\% of the responses contained fewer than 16 nodes.

Approximately 26\% of the \texttt{Neighbors} messages resulted in timeouts, exceeding the expected response window for peer communication. Geth’s default behavior is to time out after 1.5 seconds if fewer than 16 nodes are received, which can lead to incomplete node discovery and delayed network operations.

\begin{table}[H]
    \centering
    \caption{The percentage(\%) of clients sent n length of nodes to a \texttt{FindNode} request}
\vspace{-8pt}
            \begin{adjustbox}{width=\columnwidth}
    \begin{tabular}{@{}ccccccc@{}}
        \toprule
         length & Geth & Erigon & Besu & Nethermind & bor & reth \\
        \midrule
        6 & 0.67 & 0.07 & 0 & 0 & 87.28 & 0 \\ 
        9 & 1.46 & 0.22 & 0 & 0.11 & 87.05 & 0 \\ 
        10 & 1.61 & 0.11 & 0.26 & 0.41 & 90.05 & 0 \\ 
        12 & 42.22 & 8.81 & 4.21 & 16.64 & 15.51 & 0.83 \\ 
        13 & 1.92 & 0.20 & 71.89 & 0 & 0.10 & 0 \\ 
        16 & 49.87 & 11.79 & 0 & 0 & 19.80 & 0.86 \\
         \bottomrule
    \end{tabular}
    \end{adjustbox}
    \label{tab:neighbor-len-client}
\end{table}

\vspace{-12pt}

The data in Table~\ref{tab:neighbor-len-client} illustrates the variations in node counts across different clients in response to a \texttt{FindNode} request. Geth, along with Erigon and Reth (both of which are heavily based on Geth’s P2P code), consistently adheres to the specification, typically sending two \texttt{Neighbors} messages: one containing 12 nodes and another with 4 nodes, for a total of 16. This confirms these clients’ compliance with the devp2p protocol.

However, Besu deviates from this pattern by frequently responding with 13 nodes per \texttt{Neighbors} message. While this still works for most IPv4 responses, it risks exceeding the 1280-byte limit when handling IPv6 addresses, potentially causing message rejection or processing failures. Similarly, Nethermind tends to send 12 nodes, which aligns with the IPv6 limit but can still contribute to timeouts when interacting with more strict implementations like Geth.
The Bor client, which is used by the Polygon blockchain, presents another issue. Despite being a fork of Geth, Bor often sends \texttt{Neighbors} messages with significantly fewer nodes than other clients. This behavior may indicate a less robust network within Polygon’s infrastructure, which could affect the reliability of node discovery in its ecosystem.

\textbf{Potential Risks of Non-compliant Implementations.} 
Non-compliant implementations pose several risks to the Ethereum network. For instance, Besu’s tendency to send 13 nodes in response to \texttt{FindNode} requests could exceed the 1280-byte UDP limit when IPv6 addresses are used, leading to message rejection by stricter clients. This could isolate Besu clients, leaving them more vulnerable to network segmentation or targeted attacks.
On the other hand, Geth’s strict adherence to protocol specifications, requiring exactly 16 nodes or a timeout, can lead to inefficiencies. When interacting with clients like Nethermind or Besu that send fewer nodes or exceed size limits, Geth waits for the full response or times out, slowing down the node discovery process. This can impact overall network performance, especially in environments with high client diversity (see Appendix C for further technical details)

\subsection{Chain Diversity and Its Impact on Connection Efficiency}
The devp2p protocol stack supports multiple sub-protocols, such as the Ethereum mainnet protocol and the SNAP protocol, running concurrently on the same P2P overlay. Despite sharing the same node discovery protocol, these sub-protocols often have differing configuration parameters, even with the same name and version. This creates inconsistencies that can fragment the network, isolating peers with incompatible configurations.

\textbf{Diversity in Ethereum's P2P Overlay.}
As shown in Table~\ref{tab:eth-caps}, 78.05\% of peers are currently running the eth-68 mainnet sub-protocol, and 76.73\% support snap-1, which facilitates fast snapshot synchronization. Older versions like eth-66 and eth-67 remain in use, highlighting the ongoing diversity of protocols. Compared to Kim et al.~\cite{kimMeasuringEthereumNetwork2018}, non-blockchain sub-protocols have significantly decreased, reflecting Ethereum’s evolution towards a more blockchain-centric development ecosystem.
\vspace{-6pt}
\begin{table}[H]
    \centering
    \caption{Peers with sub-protocols run on the Ethereum network}\vspace{-8pt}
    \begin{tabularx}{0.4\textwidth}{XXX}
        \toprule
         Sub-protocols & Count & Percentage(\%) \\
        \midrule
        eth-68 & 53,125 & 78.05 \\
        snap-1 & 52,222 & 76.73 \\
        eth-66 & 43,075 & 63.29 \\
        eth-67 & 39,713 & 58.35 \\
        nodedata-1 & 4,800 & 7.05 \\
        eth-63 & 4,513 & 6.63 \\
        eth-65 & 4,432 & 6.51 \\
        eth-64 & 4,388 & 6.45 \\
        bsc-1 & 2,531 & 3.72 \\
        opera-63 & 2,064 & 3.03 \\
        fsnap-1 & 2,053 & 3.02 \\
         \bottomrule
    \end{tabularx}
    \label{tab:eth-caps}
\end{table}

\vspace{-6pt}

\begin{figure}[H]

    \centering
    \includegraphics[width=1\linewidth]{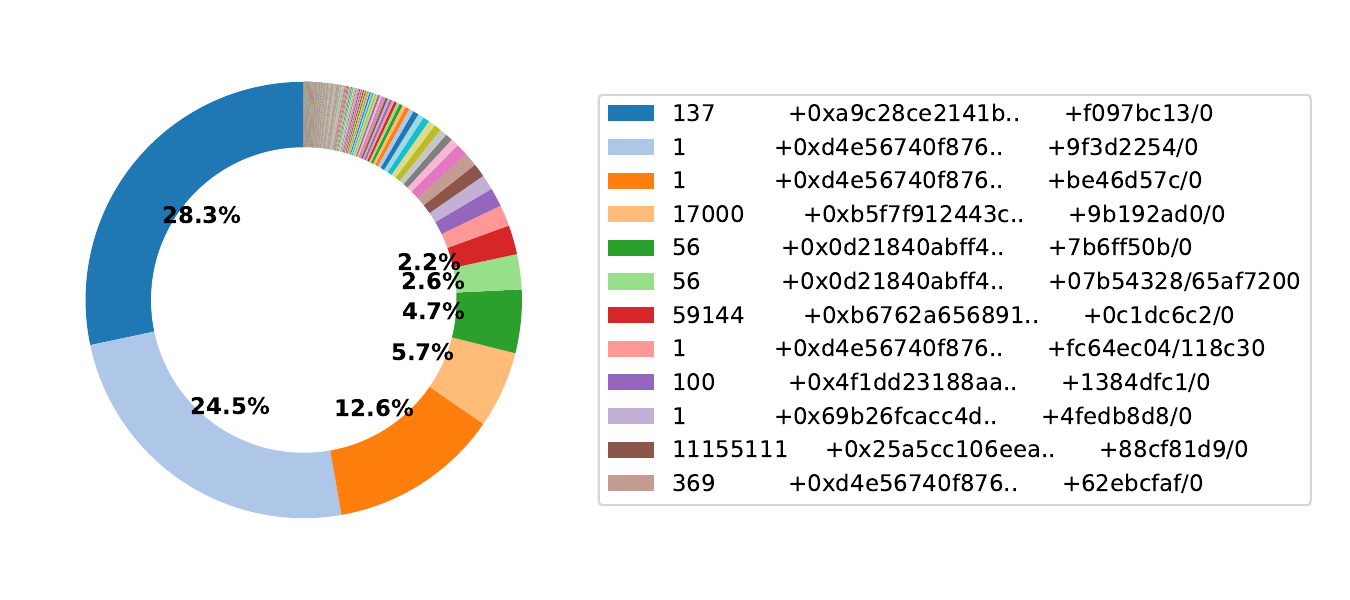}\vspace{-12pt} 
    \caption{Chain parameters distribution. Columns represent network ID, genesis, and fork ID.}
    \label{fig:chain-param}
\end{figure}

However, as seen in Figure~\ref{fig:chain-param}, only 24.5\% of peers adhere to the correct Ethereum mainnet configuration (network ID: 1, genesis: \texttt{0xd4e56740…}, fork ID: 9f3d2254/0). This marks a steep decline from the previously reported 54.5\%, meaning fewer peers now have fully compatible mainnet configurations. Consequently, only 19\% of the eth-68 peers maintain the proper mainnet settings, which drastically reduces connection efficiency.

\vspace{-10pt}
\vspace{-8pt}

Further, 12.6\% of mainnet peers are still using outdated configurations, such as those predating the Cancun upgrade (as of March 13th, 2024~\cite{ethereum_cancun}), isolating themselves from fully updated nodes. Additionally, alternative networks—such as Polygon (network ID 137), Ethereum Holesky testnet (network ID 17000), BNB Smart Chain (network ID 56), and Linea Layer 2—dominate the remaining peer configurations. Notably, Polygon’s bor clients returned \texttt{Neighbors} messages almost exclusively filled with other bor peers, suggesting that their distributed hash table is filtered to prioritize Polygon-specific peers, impacting cross-network discoverability.

\textbf{Evaluating Connection Efficiency Amidst Chain Diversity.}
The inherent chain diversity within Ethereum’s P2P network significantly impacts connection efficiency. To quantify this, we conducted a series of dialing experiments, gradually increasing the number of dial attempts and tracking peer reachability and compatibility. As detailed in Table~\ref{tab:dial-connect}, we observed that while the total number of unique peers discovered increased with more dial attempts, the proportion of fully compatible peers (Category C) sharply decreased, highlighting the challenges of maintaining efficient connections as network diversity grows.

\begin{table}[H]
    \centering
    \caption{Connection efficiency as dial attempts increase. Peers are classified as A) unreachable, B) non-compatible chain parameters, and C) fully compatible.}
    \vspace{-8pt}
    \resizebox{0.99\linewidth}{!}{
    \begin{tabular}{@{}lccccc@{}}
        \toprule
        Attempts & Unique Peers & Duplicate & A(\%) & B(\%) & C(\%) \\
        \midrule
          1,000,000    &53,685 & 19 & 29.51 & 57.42 & 13.07 \\
          5,000,000    &64,999 & 77 & 35.18 & 50.30 & 12.51 \\
          10,000,000   &74,398 & 134 & 40.61 & 47.90 & 11.49 \\
          100,000,000  &223,652& 447 &77.24&18.18 & 4.58 \\
          500,000,000  &527,957& 947 &88.96& 8.57& 2.47 \\
        \bottomrule
    \end{tabular}
    }
    \label{tab:dial-connect}
\end{table}

As dial attempts scaled from 1 million to 500 million, the number of unique peers reached increased dramatically. However, the percentage of fully compatible peers (Category C) dropped from 13.07\% to just 2.47\%, demonstrating the growing difficulty of establishing meaningful connections in a network with diverse chain parameters. The proportion of unreachable peers (Category A) surged from 29.51\% to 88.96\%, indicating that many nodes were either inactive or operating on incompatible network configurations.

Our experiments revealed that, on average, it took 947 connection attempts to successfully reach a live peer, underscoring the inefficiency caused by non-responsive nodes. These results highlight a critical issue: the Ethereum network’s distributed hash table includes many inactive or non-compatible peers, which significantly hampers the efficiency of the dial scheduler. The proliferation of nodes running different or outdated configurations introduces irrelevant node records into the network, complicating peer discovery and reducing overall connection success rates.

\textbf{Addressing Chain Diversity Challenges.}
Chain diversity, while a natural outcome of Ethereum’s expansion and evolution, introduces significant challenges in maintaining connection efficiency. Clients like bor (used by Polygon) and others running on non-mainnet networks frequently respond to FindNode requests with peers that do not meet the compatibility requirements of Ethereum’s mainnet, increasing the chances of timeouts or failed connections.

Moreover, the high percentage of unreachable nodes in our measurements reflects a common issue in decentralized systems, where peer availability fluctuates~\cite{wangBetterUnderstandingBitcoin2017}. This points to a pressing need for more intelligent distributed hash table mechanisms in Ethereum, which could prioritize peers based on recent activity and protocol compatibility, ultimately improving the network’s connectivity and resilience.

\section{Conclusion}

This study examined the Ethereum network’s P2P infrastructure, providing insights into the complexities of maintaining efficient and secure connections. By implementing a real-time analysis tool, we captured both packet-level communications and client states, offering a deeper understanding of peer interactions.

Our findings show that while Ethereum’s open protocol fosters innovation, the diversity in client configurations and sub-protocols creates challenges for scalability and efficiency. As chain diversity increases, connection inefficiencies grow, with many peers becoming unreachable or incompatible. To ensure the network’s growth and resilience, ongoing improvements in protocol design and management are essential.

\section{RELATED WORK}

Kim et al. analyzed macro-level characteristics of Ethereum’s P2P network, such as client identities and IP distributions~\cite{kimMeasuringEthereumNetwork2018}. In contrast, our research focuses on the micro-level challenges of client incompatibility and its effects on network efficiency. Gao et al. and Maeng et al. examined Ethereum’s inactive topology and node connectivity~\cite{gao2019topology, maengAnalysisEthereumNetwork2020}, while Mohammad Z. et al. reassessed peer characteristics, including geolocation and availability~\cite{masoudMeasurementStudyEthereum2024}. Tools like Ethernodes and Etherscan provide real-time views of peer statuses, and Xia et al. used message sniffing to monitor node activity~\cite{ethernode,nodetracker, xiaEthSnifferGlobalPassive2021a}.

Kiffer et al. explored the usefulness of peers in the Ethereum mainnet, showing that most connected peers offer limited protocol-level value~\cite{kifferHoodEthereumGossip2021}. Their findings align with ours, as we also identify inefficiencies in the discovery protocol that lead to few useful peer connections. Our study extends this by examining the role of client diversity and protocol incompatibilities in reducing effective connectivity, underscoring the need for more efficient discovery mechanisms.

Heo et al. examined block arrival times and DoS vulnerabilities~\cite{heoBlockExplorersPublic2021,heoPartitioningEthereumEclipsing2023}, while Kiffer et al. monitored block gossip and mining power distribution~\cite{kifferHoodEthereumGossip2021}. Li et al. introduced techniques to analyze network topology via transaction eviction rules~\cite{liTopoShotUncoveringEthereum2021}, and Zhao et al. enhanced these methods for operational efficiency~\cite{zhaoDEthnaAccurateEthereum2024}. Other studies focused on transaction pool management and DoS strategies~\cite{liDETERDenialEthereum2021,cryptoeprint:2023/956,wang2023understanding}.

\newpage
\bibliographystyle{ACM-Reference-Format}
\bibliography{refs}


\begin{thebibliography}{53}


\ifx \showCODEN    \undefined \def \showCODEN     #1{\unskip}     \fi
\ifx \showDOI      \undefined \def \showDOI       #1{#1}\fi
\ifx \showISBNx    \undefined \def \showISBNx     #1{\unskip}     \fi
\ifx \showISBNxiii \undefined \def \showISBNxiii  #1{\unskip}     \fi
\ifx \showISSN     \undefined \def \showISSN      #1{\unskip}     \fi
\ifx \showLCCN     \undefined \def \showLCCN      #1{\unskip}     \fi
\ifx \shownote     \undefined \def \shownote      #1{#1}          \fi
\ifx \showarticletitle \undefined \def \showarticletitle #1{#1}   \fi
\ifx \showURL      \undefined \def \showURL       {\relax}        \fi
\providecommand\bibfield[2]{#2}
\providecommand\bibinfo[2]{#2}
\providecommand\natexlab[1]{#1}
\providecommand\showeprint[2][]{arXiv:#2}

\bibitem[Alpha(2024)]%
        {client_diversity}
\bibfield{author}{\bibinfo{person}{Ether Alpha}.} \bibinfo{year}{2024}\natexlab{}.
\newblock \bibinfo{title}{Improve Ethereum's resilience by using a minority client}.
\newblock \bibinfo{howpublished}{\url{https://clientdiversity.org/}}.
\newblock
\newblock
\shownote{Accessed: 2024-05-11}.


\bibitem[Apriani and Sari(2021)]%
        {apriani2021performance}
\bibfield{author}{\bibinfo{person}{Mega Apriani} {and} \bibinfo{person}{Riri~Fitri Sari}.} \bibinfo{year}{2021}\natexlab{}.
\newblock \showarticletitle{Performance Comparison of Spongent and Photon Hashing Algorithms in Ethereum-based Blockchain System}. In \bibinfo{booktitle}{\emph{2021 7th International Conference on Electrical, Electronics and Information Engineering (ICEEIE)}}. \bibinfo{pages}{564--569}.
\newblock
\urldef\tempurl%
\url{https://doi.org/10.1109/ICEEIE52663.2021.9616831}
\showDOI{\tempurl}


\bibitem[ChainList(2021)]%
        {chainlist}
\bibfield{author}{\bibinfo{person}{ChainList}.} \bibinfo{year}{2021}\natexlab{}.
\newblock \bibinfo{title}{Helping users connect to EVM powered networks}.
\newblock \bibinfo{howpublished}{\url{https://chainlist.org/}}.
\newblock
\newblock
\shownote{Accessed: 2024-04-10}.


\bibitem[Dan~Cline(2024)]%
        {reth_pull_4849}
\bibfield{author}{\bibinfo{person}{Matthias~Seitz1 Dan~Cline}.} \bibinfo{year}{2024}\natexlab{}.
\newblock \bibinfo{title}{Add an ability to change P2P listening address}.
\newblock \bibinfo{howpublished}{\url{https://github.com/paradigmxyz/reth/pull/4849}}.
\newblock
\newblock
\shownote{Accessed: 2024-04-10}.


\bibitem[{etclabscore}(NA)]%
        {github-core-geth}
\bibfield{author}{\bibinfo{person}{{etclabscore}}.} \bibinfo{year}{N/A}\natexlab{}.
\newblock \bibinfo{title}{{GitHub - etclabscore/core-geth: A highly configurable Go implementation of the Ethereum protocol}}.
\newblock
\newblock
\urldef\tempurl%
\url{https://github.com/etclabscore/core-geth}
\showURL{%
\tempurl}
\newblock
\shownote{GitHub repository}.


\bibitem[{ETCMC DAO}(2024)]%
        {etcmc}
\bibfield{author}{\bibinfo{person}{{ETCMC DAO}}.} \bibinfo{year}{2024}\natexlab{}.
\newblock \bibinfo{title}{{ETCMC - Decentralise the Ethereum Classic Network}}.
\newblock \bibinfo{howpublished}{\url{https://etcmc.org/}}.
\newblock
\newblock
\shownote{Accessed on May 6, 2024}.


\bibitem[{Ethereum}(2015)]%
        {ethereum_smart_contracts}
\bibfield{author}{\bibinfo{person}{{Ethereum}}.} \bibinfo{year}{2015}\natexlab{}.
\newblock \bibinfo{title}{Introduction to Smart Contracts}.
\newblock \bibinfo{howpublished}{\url{https://ethereum.org/en/developers/docs/smart-contracts/}}.
\newblock
\newblock
\shownote{Accessed: 2024-04-10}.


\bibitem[Ethereum(2018)]%
        {ethereum_whisper}
\bibfield{author}{\bibinfo{person}{Ethereum}.} \bibinfo{year}{2018}\natexlab{}.
\newblock \bibinfo{title}{Whisper: Ethereum Communication Protocol}.
\newblock \bibinfo{howpublished}{\url{https://github.com/ethereum/whisper}}.
\newblock
\newblock
\shownote{Accessed: 2024-04-10}.


\bibitem[Ethereum(2024a)]%
        {ethereum_devp2p}
\bibfield{author}{\bibinfo{person}{Ethereum}.} \bibinfo{year}{2024}\natexlab{a}.
\newblock \bibinfo{title}{devp2p: Ethereum peer-to-peer networking specifications}.
\newblock \bibinfo{howpublished}{\url{https://github.com/ethereum/devp2p}}.
\newblock
\newblock
\shownote{Accessed: 2024-04-10}.


\bibitem[Ethereum(2024b)]%
        {discv4_findnode}
\bibfield{author}{\bibinfo{person}{Ethereum}.} \bibinfo{year}{2024}\natexlab{b}.
\newblock \bibinfo{title}{Finnode message in Ethereum Node Discovery Protocol v4}.
\newblock \bibinfo{howpublished}{\url{https://github.com/ethereum/devp2p/blob/master/discv4.md\#findnode-packet-0x03}}.
\newblock
\newblock
\shownote{Accessed: 2024-04-30}.


\bibitem[Ethereum(2024c)]%
        {go_ethereum}
\bibfield{author}{\bibinfo{person}{Ethereum}.} \bibinfo{year}{2024}\natexlab{c}.
\newblock \bibinfo{title}{go-ethereum: Official Go implementation of the Ethereum protocol}.
\newblock \bibinfo{howpublished}{\url{https://github.com/ethereum/go-ethereum}}.
\newblock
\newblock
\shownote{Accessed: 2024-04-10}.


\bibitem[{Ethereum Execution Specs}(2023)]%
        {ethereum_paris_upgrade}
\bibfield{author}{\bibinfo{person}{{Ethereum Execution Specs}}.} \bibinfo{year}{2023}\natexlab{}.
\newblock \bibinfo{title}{Mainnet Upgrade: Paris}.
\newblock \bibinfo{howpublished}{\url{https://github.com/ethereum/execution-specs/blob/master/network-upgrades/mainnet-upgrades/paris.md}}.
\newblock


\bibitem[{Ethereum Execution Specs}(2024)]%
        {ethereum_cancun}
\bibfield{author}{\bibinfo{person}{{Ethereum Execution Specs}}.} \bibinfo{year}{2024}\natexlab{}.
\newblock \bibinfo{title}{Cancun Network Upgrade Specification}.
\newblock \bibinfo{howpublished}{\url{https://github.com/ethereum/execution-specs/blob/master/network-upgrades/mainnet-upgrades/cancun.md}}.
\newblock


\bibitem[{Ethereum Foundation}(2023)]%
        {go_ethereum_lookup_2023}
\bibfield{author}{\bibinfo{person}{{Ethereum Foundation}}.} \bibinfo{year}{2023}\natexlab{}.
\newblock \bibinfo{title}{go-ethereum: {P2P} Discovery Lookup Code}.
\newblock \bibinfo{howpublished}{\url{https://github.com/ethereum/go-ethereum/blob/65e5ca7d8126f7a8c708f8affb64f16c22cc63c0/p2p/discover/lookup.go\#L142-L152}}.
\newblock
\newblock
\shownote{Accessed: 2024-10-05}.


\bibitem[{Ethereum Foundation}(2024a)]%
        {ethereum_execution_specs}
\bibfield{author}{\bibinfo{person}{{Ethereum Foundation}}.} \bibinfo{year}{2024}\natexlab{a}.
\newblock \bibinfo{title}{Ethereum Execution Client Specifications}.
\newblock \bibinfo{howpublished}{\url{https://github.com/ethereum/execution-specs}}.
\newblock
\newblock
\shownote{Accessed: 2024-04-10}.


\bibitem[{Ethereum Foundation}(2024b)]%
        {ethereum_snap_protocol_2024}
\bibfield{author}{\bibinfo{person}{{Ethereum Foundation}}.} \bibinfo{year}{2024}\natexlab{b}.
\newblock \bibinfo{title}{Ethereum Snapshot Protocol (SNAP)}.
\newblock \bibinfo{howpublished}{\url{https://github.com/ethereum/devp2p/blob/master/caps/snap.md}}.
\newblock
\newblock
\shownote{Accessed: 2024-10-05}.


\bibitem[{Ethereum Foundation}(2024c)]%
        {ethereum_eth_protocol_2024}
\bibfield{author}{\bibinfo{person}{{Ethereum Foundation}}.} \bibinfo{year}{2024}\natexlab{c}.
\newblock \bibinfo{title}{Ethereum Wire Protocol}.
\newblock \bibinfo{howpublished}{\url{https://github.com/ethereum/devp2p/blob/master/caps/eth.md}}.
\newblock
\newblock
\shownote{Accessed: 2024-10-05}.


\bibitem[Ethernodes(2024)]%
        {ethernode}
\bibfield{author}{\bibinfo{person}{Ethernodes}.} \bibinfo{year}{2024}\natexlab{}.
\newblock \bibinfo{title}{Ethereum mainnet statistics}.
\newblock \bibinfo{howpublished}{\url{https://ethernodes.org/}}.
\newblock
\newblock
\shownote{Accessed: 2024-05-14}.


\bibitem[Etherscan(2024)]%
        {nodetracker}
\bibfield{author}{\bibinfo{person}{Etherscan}.} \bibinfo{year}{2024}\natexlab{}.
\newblock \bibinfo{title}{Ethereum node tracker}.
\newblock \bibinfo{howpublished}{\url{https://etherscan.io/nodetracker}}.
\newblock
\newblock
\shownote{Accessed: 2024-05-14}.


\bibitem[Foundation(2022)]%
        {ethereum_merge_roadmap}
\bibfield{author}{\bibinfo{person}{Ethereum Foundation}.} \bibinfo{year}{2022}\natexlab{}.
\newblock \bibinfo{title}{The Merge}.
\newblock \bibinfo{howpublished}{\url{https://ethereum.org/en/roadmap/merge/}}.
\newblock


\bibitem[Foundation(2019)]%
        {swarm}
\bibfield{author}{\bibinfo{person}{Swarm Foundation}.} \bibinfo{year}{2019}\natexlab{}.
\newblock \bibinfo{title}{Swarm: Decentralised Data Storage and Distribution Technology}.
\newblock \bibinfo{howpublished}{\url{https://www.ethswarm.org/}}.
\newblock
\newblock
\shownote{Accessed: 2024-04-10}.


\bibitem[Gao et~al\mbox{.}(2019)]%
        {gao2019topology}
\bibfield{author}{\bibinfo{person}{Yue Gao}, \bibinfo{person}{Jinqiao Shi}, \bibinfo{person}{Xuebin Wang}, \bibinfo{person}{Qingfeng Tan}, \bibinfo{person}{Can Zhao}, {and} \bibinfo{person}{Zelin Yin}.} \bibinfo{year}{2019}\natexlab{}.
\newblock \showarticletitle{Topology measurement and analysis on ethereum p2p network}. In \bibinfo{booktitle}{\emph{2019 IEEE Symposium on Computers and Communications (ISCC)}}. IEEE, \bibinfo{pages}{1--7}.
\newblock


\bibitem[Heimbach et~al\mbox{.}(2023)]%
        {heimbachEthereumProposerBuilderSeparation2023}
\bibfield{author}{\bibinfo{person}{Lioba Heimbach}, \bibinfo{person}{Lucianna Kiffer}, \bibinfo{person}{Christof~Ferreira Torres}, {and} \bibinfo{person}{Roger Wattenhofer}.} \bibinfo{year}{2023}\natexlab{}.
\newblock \bibinfo{title}{Ethereum's Proposer-Builder Separation: Promises and Realities}.
\newblock
\newblock
\showeprint[arxiv]{2305.19037}~[cs]


\bibitem[Henningsen et~al\mbox{.}(2019)]%
        {henningsen2019eclipsing}
\bibfield{author}{\bibinfo{person}{Sebastian Henningsen}, \bibinfo{person}{Daniel Teunis}, \bibinfo{person}{Martin Florian}, {and} \bibinfo{person}{Bj{\"o}rn Scheuermann}.} \bibinfo{year}{2019}\natexlab{}.
\newblock \showarticletitle{Eclipsing ethereum peers with false friends}.
\newblock \bibinfo{journal}{\emph{arXiv preprint arXiv:1908.10141}} (\bibinfo{year}{2019}).
\newblock


\bibitem[Heo and Shin(2021)]%
        {heoBlockExplorersPublic2021}
\bibfield{author}{\bibinfo{person}{Hwanjo Heo} {and} \bibinfo{person}{Seungwon Shin}.} \bibinfo{year}{2021}\natexlab{}.
\newblock \showarticletitle{Behind Block Explorers: Public Blockchain Measurement and Security Implication}. In \bibinfo{booktitle}{\emph{2021 IEEE 41st International Conference on Distributed Computing Systems (ICDCS)}}. \bibinfo{publisher}{IEEE}, \bibinfo{address}{DC, USA}, \bibinfo{pages}{216--226}.
\newblock
\showISBNx{978-1-66544-513-9}
\urldef\tempurl%
\url{https://doi.org/10.1109/ICDCS51616.2021.00029}
\showDOI{\tempurl}


\bibitem[Heo et~al\mbox{.}(2023)]%
        {heoPartitioningEthereumEclipsing2023}
\bibfield{author}{\bibinfo{person}{Hwanjo Heo}, \bibinfo{person}{Seungwon Woo}, \bibinfo{person}{Taeung Yoon}, \bibinfo{person}{Min~Suk Kang}, {and} \bibinfo{person}{Seungwon Shin}.} \bibinfo{year}{2023}\natexlab{}.
\newblock \showarticletitle{Partitioning Ethereum without Eclipsing It}. In \bibinfo{booktitle}{\emph{Proceedings 2023 Network and Distributed System Security Symposium}}. \bibinfo{publisher}{Internet Society}, \bibinfo{address}{San Diego, CA, USA}.
\newblock
\showISBNx{978-1-891562-83-9}
\urldef\tempurl%
\url{https://doi.org/10.14722/ndss.2023.24465}
\showDOI{\tempurl}


\bibitem[Johnson et~al\mbox{.}(2001)]%
        {johnson2001elliptic}
\bibfield{author}{\bibinfo{person}{Don Johnson}, \bibinfo{person}{Alfred Menezes}, {and} \bibinfo{person}{Scott Vanstone}.} \bibinfo{year}{2001}\natexlab{}.
\newblock \showarticletitle{The elliptic curve digital signature algorithm (ECDSA)}.
\newblock \bibinfo{journal}{\emph{International journal of information security}}  \bibinfo{volume}{1} (\bibinfo{year}{2001}), \bibinfo{pages}{36--63}.
\newblock


\bibitem[Kempton(2024)]%
        {eip7636}
\bibfield{author}{\bibinfo{person}{James Kempton}.} \bibinfo{year}{2024}\natexlab{}.
\newblock \bibinfo{title}{EIP-7636: Extension of EIP-778 for "client" ENR Entry [DRAFT]}.
\newblock \bibinfo{howpublished}{\url{https://eips.ethereum.org/EIPS/eip-7636}}.
\newblock
\newblock
\shownote{[Online serial]}.


\bibitem[Kiffer et~al\mbox{.}(2021)]%
        {kifferHoodEthereumGossip2021}
\bibfield{author}{\bibinfo{person}{Lucianna Kiffer}, \bibinfo{person}{Asad Salman}, \bibinfo{person}{Dave Levin}, \bibinfo{person}{Alan Mislove}, {and} \bibinfo{person}{Cristina {Nita-Rotaru}}.} \bibinfo{year}{2021}\natexlab{}.
\newblock \showarticletitle{Under the Hood of the Ethereum Gossip Protocol}.
\newblock In \bibinfo{booktitle}{\emph{Financial Cryptography and Data Security}}, \bibfield{editor}{\bibinfo{person}{Nikita Borisov} {and} \bibinfo{person}{Claudia Diaz}} (Eds.). Vol.~\bibinfo{volume}{12675}. \bibinfo{publisher}{Springer Berlin Heidelberg}, \bibinfo{address}{Berlin, Heidelberg}, \bibinfo{pages}{437--456}.
\newblock
\showISBNx{978-3-662-64330-3 978-3-662-64331-0}
\urldef\tempurl%
\url{https://doi.org/10.1007/978-3-662-64331-0_23}
\showDOI{\tempurl}


\bibitem[Kim et~al\mbox{.}(2018)]%
        {kimMeasuringEthereumNetwork2018}
\bibfield{author}{\bibinfo{person}{Seoung~Kyun Kim}, \bibinfo{person}{Zane Ma}, \bibinfo{person}{Siddharth Murali}, \bibinfo{person}{Joshua Mason}, \bibinfo{person}{Andrew Miller}, {and} \bibinfo{person}{Michael Bailey}.} \bibinfo{year}{2018}\natexlab{}.
\newblock \showarticletitle{Measuring Ethereum Network Peers}. In \bibinfo{booktitle}{\emph{Proceedings of the Internet Measurement Conference 2018}}. \bibinfo{publisher}{ACM}, \bibinfo{address}{Boston MA USA}, \bibinfo{pages}{91--104}.
\newblock
\showISBNx{978-1-4503-5619-0}
\urldef\tempurl%
\url{https://doi.org/10.1145/3278532.3278542}
\showDOI{\tempurl}


\bibitem[Li et~al\mbox{.}(2021a)]%
        {liTopoShotUncoveringEthereum2021}
\bibfield{author}{\bibinfo{person}{Kai Li}, \bibinfo{person}{Yuzhe Tang}, \bibinfo{person}{Jiaqi Chen}, \bibinfo{person}{Yibo Wang}, {and} \bibinfo{person}{Xianghong Liu}.} \bibinfo{year}{2021}\natexlab{a}.
\newblock \showarticletitle{TopoShot: Uncovering Ethereum's Network Topology Leveraging Replacement Transactions}. In \bibinfo{booktitle}{\emph{Proceedings of the 21st ACM Internet Measurement Conference}}. \bibinfo{publisher}{ACM}, \bibinfo{address}{Virtual Event}, \bibinfo{pages}{302--319}.
\newblock
\showISBNx{978-1-4503-9129-0}
\urldef\tempurl%
\url{https://doi.org/10.1145/3487552.3487814}
\showDOI{\tempurl}


\bibitem[Li et~al\mbox{.}(2021b)]%
        {liDETERDenialEthereum2021}
\bibfield{author}{\bibinfo{person}{Kai Li}, \bibinfo{person}{Yibo Wang}, {and} \bibinfo{person}{Yuzhe Tang}.} \bibinfo{year}{2021}\natexlab{b}.
\newblock \showarticletitle{DETER: Denial of Ethereum Txpool sERvices}. In \bibinfo{booktitle}{\emph{Proceedings of the 2021 ACM SIGSAC Conference on Computer and Communications Security}} \emph{(\bibinfo{series}{CCS '21})}. \bibinfo{publisher}{Association for Computing Machinery}, \bibinfo{address}{New York, NY, USA}, \bibinfo{pages}{1645--1667}.
\newblock
\showISBNx{978-1-4503-8454-4}
\urldef\tempurl%
\url{https://doi.org/10.1145/3460120.3485369}
\showDOI{\tempurl}


\bibitem[Maeng et~al\mbox{.}(2020)]%
        {maengAnalysisEthereumNetwork2020}
\bibfield{author}{\bibinfo{person}{Soo~Hoon Maeng}, \bibinfo{person}{Meryam Essaid}, {and} \bibinfo{person}{Hong~Taek Ju}.} \bibinfo{year}{2020}\natexlab{}.
\newblock \showarticletitle{Analysis of Ethereum Network Properties and Behavior of Influential Nodes}. In \bibinfo{booktitle}{\emph{2020 21st Asia-Pacific Network Operations and Management Symposium (APNOMS)}}. \bibinfo{pages}{203--207}.
\newblock
\showISSN{2576-8565}
\urldef\tempurl%
\url{https://doi.org/10.23919/APNOMS50412.2020.9236965}
\showDOI{\tempurl}


\bibitem[Masoud et~al\mbox{.}(2024)]%
        {masoudMeasurementStudyEthereum2024}
\bibfield{author}{\bibinfo{person}{Mohammad~Z. Masoud}, \bibinfo{person}{Yousef Jaradat}, \bibinfo{person}{Ahmad Manasrah}, \bibinfo{person}{Mohammad Alia}, \bibinfo{person}{Khaled Suwais}, {and} \bibinfo{person}{Sally Almanasra}.} \bibinfo{year}{2024}\natexlab{}.
\newblock \showarticletitle{A Measurement Study of the Ethereum Underlying P2P Network}.
\newblock \bibinfo{journal}{\emph{Computers, Materials \& Continua}} \bibinfo{volume}{78}, \bibinfo{number}{1} (\bibinfo{year}{2024}), \bibinfo{pages}{515--532}.
\newblock
\showISSN{1546-2226}
\urldef\tempurl%
\url{https://doi.org/10.32604/cmc.2023.044504}
\showDOI{\tempurl}


\bibitem[{Matic Network}(2024)]%
        {polygon_bor}
\bibfield{author}{\bibinfo{person}{{Matic Network}}.} \bibinfo{year}{2024}\natexlab{}.
\newblock \bibinfo{title}{Official repository for the Polygon Blockchain}.
\newblock \bibinfo{howpublished}{\url{https://github.com/maticnetwork/bor}}.
\newblock
\newblock
\shownote{Accessed: 2024-04-10}.


\bibitem[Maymounkov and Mazieres(2002)]%
        {maymounkov2002kademlia}
\bibfield{author}{\bibinfo{person}{Petar Maymounkov} {and} \bibinfo{person}{David Mazieres}.} \bibinfo{year}{2002}\natexlab{}.
\newblock \showarticletitle{Kademlia: A peer-to-peer information system based on the xor metric}. In \bibinfo{booktitle}{\emph{International Workshop on Peer-to-Peer Systems}}. Springer, \bibinfo{pages}{53--65}.
\newblock


\bibitem[Seitz(2024)]%
        {reth_issue_4222}
\bibfield{author}{\bibinfo{person}{Matthias Seitz}.} \bibinfo{year}{2024}\natexlab{}.
\newblock \bibinfo{title}{Support external address in discv4 handle type}.
\newblock \bibinfo{howpublished}{\url{https://github.com/paradigmxyz/reth/issues/4222}}.
\newblock
\newblock
\shownote{Accessed: 2024-04-10}.


\bibitem[Seitz1(2024)]%
        {reth_pull_4224}
\bibfield{author}{\bibinfo{person}{Matthias Seitz1}.} \bibinfo{year}{2024}\natexlab{}.
\newblock \bibinfo{title}{Feature: Track Node Record}.
\newblock \bibinfo{howpublished}{\url{https://github.com/paradigmxyz/reth/pull/4224}}.
\newblock
\newblock
\shownote{Accessed: 2024-04-10}.


\bibitem[Seitz2(2024)]%
        {reth_pull_4268}
\bibfield{author}{\bibinfo{person}{Matthias Seitz2}.} \bibinfo{year}{2024}\natexlab{}.
\newblock \bibinfo{title}{Fix: Prevent Node Info Zero Address}.
\newblock \bibinfo{howpublished}{\url{https://github.com/paradigmxyz/reth/pull/4268}}.
\newblock
\newblock
\shownote{Accessed: 2024-04-10}.


\bibitem[Szilágyi(2024)]%
        {geth_29329}
\bibfield{author}{\bibinfo{person}{Péter Szilágyi}.} \bibinfo{year}{2024}\natexlab{}.
\newblock \bibinfo{title}{Connection Slot Exhaustion with Passive Nodes}.
\newblock \bibinfo{howpublished}{\url{https://github.com/ethereum/go-ethereum/issues/29329\#issuecomment-2031290591}}.
\newblock
\newblock
\shownote{Accessed: 2024-05-08}.


\bibitem[Szilágyi and Lange(2019)]%
        {eip2124}
\bibfield{author}{\bibinfo{person}{Péter Szilágyi} {and} \bibinfo{person}{Felix Lange}.} \bibinfo{year}{2019}\natexlab{}.
\newblock \bibinfo{title}{EIP-2124: Fork identifier for chain compatibility checks}.
\newblock \bibinfo{howpublished}{\url{https://eips.ethereum.org/EIPS/eip-2124}}.
\newblock
\newblock
\shownote{[Online serial]}.


\bibitem[Torres et~al\mbox{.}(2021)]%
        {torres2021frontrunner}
\bibfield{author}{\bibinfo{person}{Christof~Ferreira Torres}, \bibinfo{person}{Ramiro Camino}, {and} \bibinfo{person}{Radu State}.} \bibinfo{year}{2021}\natexlab{}.
\newblock \showarticletitle{Frontrunner Jones and the Raiders of the Dark Forest: An Empirical Study of Frontrunning on the Ethereum Blockchain}. In \bibinfo{booktitle}{\emph{30th USENIX Security Symposium (USENIX Security 21)}}. \bibinfo{publisher}{USENIX Association}, \bibinfo{pages}{1343--1359}.
\newblock
\showISBNx{978-1-939133-24-3}
\urldef\tempurl%
\url{https://www.usenix.org/conference/usenixsecurity21/presentation/torres}
\showURL{%
\tempurl}


\bibitem[van~der Wijden(2022)]%
        {eip6122}
\bibfield{author}{\bibinfo{person}{Marius van~der Wijden}.} \bibinfo{year}{2022}\natexlab{}.
\newblock \bibinfo{title}{EIP-6122: Forkid checks based on timestamps}.
\newblock \bibinfo{howpublished}{\url{https://eips.ethereum.org/EIPS/eip-6122}}.
\newblock
\newblock
\shownote{[Online serial]}.


\bibitem[Wang and Pustogarov(2017)]%
        {wangBetterUnderstandingBitcoin2017}
\bibfield{author}{\bibinfo{person}{Liang Wang} {and} \bibinfo{person}{Ivan Pustogarov}.} \bibinfo{year}{2017}\natexlab{}.
\newblock \bibinfo{title}{Towards Better Understanding of Bitcoin Unreachable Peers}.
\newblock
\newblock
\showeprint[arxiv]{1709.06837}~[cs]


\bibitem[Wang et~al\mbox{.}(2023)]%
        {wang2023understanding}
\bibfield{author}{\bibinfo{person}{Yibo Wang}, \bibinfo{person}{Wanning Ding}, \bibinfo{person}{Kai Li}, {and} \bibinfo{person}{Yuzhe Tang}.} \bibinfo{year}{2023}\natexlab{}.
\newblock \showarticletitle{Understanding ethereum mempool security under asymmetric dos by symbolic fuzzing}.
\newblock \bibinfo{journal}{\emph{arXiv preprint arXiv:2312.02642}} (\bibinfo{year}{2023}).
\newblock


\bibitem[Weintraub et~al\mbox{.}(2022)]%
        {weintraubFlashBotPan2022}
\bibfield{author}{\bibinfo{person}{Ben Weintraub}, \bibinfo{person}{Christof~Ferreira Torres}, \bibinfo{person}{Cristina {Nita-Rotaru}}, {and} \bibinfo{person}{Radu State}.} \bibinfo{year}{2022}\natexlab{}.
\newblock \showarticletitle{A Flash(Bot) in the Pan: Measuring Maximal Extractable Value in Private Pools}. In \bibinfo{booktitle}{\emph{Proceedings of the 22nd ACM Internet Measurement Conference}}. \bibinfo{publisher}{ACM}, \bibinfo{address}{Nice France}, \bibinfo{pages}{458--471}.
\newblock
\showISBNx{978-1-4503-9259-4}
\urldef\tempurl%
\url{https://doi.org/10.1145/3517745.3561448}
\showDOI{\tempurl}


\bibitem[Whitehead(2024)]%
        {besuPR6225}
\bibfield{author}{\bibinfo{person}{Matt Whitehead}.} \bibinfo{year}{2024}\natexlab{}.
\newblock \bibinfo{title}{Use From field in a PING packet when creating a peer table entry}.
\newblock \bibinfo{howpublished}{\url{https://github.com/hyperledger/besu/pull/6225}}.
\newblock
\newblock
\shownote{Accessed: 2024-05-6}.


\bibitem[Whitehead1(2024)]%
        {besuPR6439}
\bibfield{author}{\bibinfo{person}{Matt Whitehead1}.} \bibinfo{year}{2024}\natexlab{}.
\newblock \bibinfo{title}{Only accept an address from a peer if it is a valid IP address}.
\newblock \bibinfo{howpublished}{\url{https://github.com/hyperledger/besu/pull/6439}}.
\newblock
\newblock
\shownote{Accessed: 2024-05-6}.


\bibitem[Wood et~al\mbox{.}(2024)]%
        {ethereum_yellowpaper}
\bibfield{author}{\bibinfo{person}{Gavin Wood} {et~al\mbox{.}}} \bibinfo{year}{2024}\natexlab{}.
\newblock \bibinfo{title}{Ethereum: A Secure Decentralised Generalised Transaction Ledger}.
\newblock \bibinfo{howpublished}{\url{https://github.com/ethereum/yellowpaper}}.
\newblock


\bibitem[Xia et~al\mbox{.}(2021)]%
        {xiaEthSnifferGlobalPassive2021a}
\bibfield{author}{\bibinfo{person}{Wei Xia}, \bibinfo{person}{Zhenzhen Li}, \bibinfo{person}{Zhen Li}, \bibinfo{person}{Gang Xiong}, {and} \bibinfo{person}{Gaopeng Gou}.} \bibinfo{year}{2021}\natexlab{}.
\newblock \showarticletitle{EthSniffer: A Global Passive Perspective on Ethereum}.
\newblock In \bibinfo{booktitle}{\emph{Blockchain and Trustworthy Systems}}, \bibfield{editor}{\bibinfo{person}{Hong-Ning Dai}, \bibinfo{person}{Xuanzhe Liu}, \bibinfo{person}{Daniel~Xiapu Luo}, \bibinfo{person}{Jiang Xiao}, {and} \bibinfo{person}{Xiangping Chen}} (Eds.). Vol.~\bibinfo{volume}{1490}. \bibinfo{publisher}{Springer Singapore}, \bibinfo{address}{Singapore}, \bibinfo{pages}{70--84}.
\newblock
\showISBNx{9789811679926 9789811679933}
\urldef\tempurl%
\url{https://doi.org/10.1007/978-981-16-7993-3_6}
\showDOI{\tempurl}


\bibitem[Yaish et~al\mbox{.}(2023)]%
        {cryptoeprint:2023/956}
\bibfield{author}{\bibinfo{person}{Aviv Yaish}, \bibinfo{person}{Kaihua Qin}, \bibinfo{person}{Liyi Zhou}, \bibinfo{person}{Aviv Zohar}, {and} \bibinfo{person}{Arthur Gervais}.} \bibinfo{year}{2023}\natexlab{}.
\newblock \bibinfo{title}{Speculative Denial-of-Service Attacks in Ethereum}.
\newblock \bibinfo{howpublished}{Cryptology ePrint Archive, Paper 2023/956}.
\newblock
\urldef\tempurl%
\url{https://eprint.iacr.org/2023/956}
\showURL{%
\tempurl}
\newblock
\shownote{\url{https://eprint.iacr.org/2023/956}}.


\bibitem[Yi et~al\mbox{.}(2022)]%
        {yi2022blockscope}
\bibfield{author}{\bibinfo{person}{Xiao Yi}, \bibinfo{person}{Yuzhou Fang}, \bibinfo{person}{Daoyuan Wu}, {and} \bibinfo{person}{Lingxiao Jiang}.} \bibinfo{year}{2022}\natexlab{}.
\newblock \showarticletitle{BlockScope: Detecting and Investigating Propagated Vulnerabilities in Forked Blockchain Projects}.
\newblock \bibinfo{journal}{\emph{arXiv preprint arXiv:2208.00205}} (\bibinfo{year}{2022}).
\newblock


\bibitem[Zhao et~al\mbox{.}(2024)]%
        {zhaoDEthnaAccurateEthereum2024}
\bibfield{author}{\bibinfo{person}{Chonghe Zhao}, \bibinfo{person}{Yipeng Zhou}, \bibinfo{person}{Shengli Zhang}, \bibinfo{person}{Taotao Wang}, \bibinfo{person}{Quan~Z. Sheng}, {and} \bibinfo{person}{Song Guo}.} \bibinfo{year}{2024}\natexlab{}.
\newblock \bibinfo{title}{DEthna: Accurate Ethereum Network Topology Discovery with Marked Transactions}.
\newblock
\newblock
\showeprint[arxiv]{2402.03881}~[cs]


\end{thebibliography}

\appendix

\section{Client Configuration}

We optimized several parameters and code modifications to improve node performance and enhance insights into the Ethereum network, focusing on node discovery efficiency and bandwidth management.

\textbf{Node Discovery Protocol.} 
We increased the maximum nodes per bucket from 16 to 1,048,576, allowing a broader node table to enhance discovery. To prevent traffic amplification, the number of nodes sent in \texttt{Neighbors} messages remains capped at 16. We also increased concurrent node discovery tasks from 3 to 1,000, reducing the time to connect to 5,000 nodes from four days to 10 hours. Storing 1,000 nodes in the peer database for rapid reconnection further improved startup efficiency.

\textbf{DHT Management and Dial Scheduler.} 
The DHT table refresh interval was reduced from five minutes to 10 seconds to increase node responsiveness, although this raises the potential attack surface. The dial scheduler was also optimized, increasing the number of sources from 1 to 50, enabling around 100 dials per second per node.

\textbf{Ethereum Wire Protocol.} 
To reduce bandwidth consumption during transaction broadcasts, we configured the client to only announce transaction hashes, not full transactions. This prevents redundant data transmission. Our system replies to only half of the transaction requests it receives, conserving bandwidth further.

\textbf{Custom Logging and Monitoring.} 
We implemented customized logging for dial attempts and node interactions, providing detailed insights into connection failures and system performance, which is essential for understanding the network’s behavior.

We deployed five Geth full-node instances (v1.14.0) on Ubuntu 22.04 with 1 Gb/s bandwidth, 256 GB RAM, and 64 cores. This multi-node setup improved peer discovery and ensured representative data coverage of the Ethereum network.

 \section{System Configuration}
We deployed our measurement infrastructure on Ubuntu 22.04, optimizing system settings for enhanced performance, including TCP buffering, log processing, and database management.

To handle high traffic and connect with most public peers, we increased the TCP buffer size: 
1) \texttt{\seqsplit{net.core.rmem\_max / net.core.wmem\_max=16777216}}: Increased maximum receive and write buffer sizes for handling high-speed data, reducing delays and avoiding buffer overflows.
2) \texttt{\seqsplit{net.ipv4.tcp\_rmem} / \texttt{tcp\_wmem = 4096 87380 16777216}}: Adjusted minimum, default, and maximum TCP buffer sizes to accommodate high throughput environments.

\textbf{Log Processing Optimization.} The system generated significant logs from Geth’s built-in trace logs and custom outputs. We optimized log handling by: 1) Using Ubuntu’s \texttt{logger} with custom rules and \texttt{rsyslog} for efficient log rotation; 2) Automating log compression and rotation with \texttt{cron} to manage high throughput without overwhelming file descriptors.

\textbf{Database Configuration.} We utilized MongoDB for its high concurrency and schema-less document storage capabilities, optimizing it as follows:
1) Memory Usage: Minimized indexing during the collection phase to reduce memory overhead, keeping only essential indexes for real-time queries. For large datasets, we used split processing to manage oversized collections.
2) Memory Limit: Reduced MongoDB’s memory usage to 64 GB to ensure system stability.
3) Real-time Analysis: Enabled real-time data analysis to reduce the need for post-collection processing and to monitor network health efficiently.
\section{Potential Risk from Non-compliant Implementations}

\begin{enumerate}[leftmargin=*,topsep=1pt]
\item \textbf{Timeout Dependency:} When Geth sends a \texttt{FindNode} message to clients like Nethermind or Besu, it must wait for the full 1.5-second timeout before processing the nodes received. These clients typically return only 12 or 13 nodes in a \texttt{Neighbors} message, falling short of Geth’s expectation of 16 nodes and triggering the timeout.

\item \textbf{Efficiency Impacts:} The enforced wait for additional nodes or timeout significantly impacts efficiency, slowing peer discovery and dialing tasks. Unlike Geth, which waits until the timeout to process nodes, clients like Nethermind handle nodes immediately, leading to faster processing.

\item \textbf{Concurrent Lookup Limitations:} Geth limits itself to three concurrent \texttt{FindNode} tasks per lookup operation, with only one task feeding peers into the dial scheduler. If Geth queries multiple Nethermind nodes simultaneously, cumulative delays from waiting for responses can further slow down the discovery process.

\item \textbf{Dial Selection Delays:} Geth depends on its lookup buffer to select peers for dialing, with only one random-target task running at any given time. Prolonged discovery phases delay the dialing of new peers, degrading overall network performance.

\end{enumerate}

These constraints highlight the need for protocol adjustments or optimizations in Geth’s handling of \texttt{FindNode} responses to improve network efficiency. The devp2p protocol stack’s specifications are not as strict or well-defined as the Ethereum Virtual Machine (EVM) specs, often relying on developer coordination~\cite{ethereum_execution_specs}. While this flexibility mitigates conflicts, new challenges, such as IPv6 adoption and updated peer scoring mechanisms, introduce additional risks. These evolving scenarios necessitate a reevaluation of the protocol to enhance stability and security across the network.

\section{Interaction between Peers}
In Ethereum’s node discovery process, peer-to-peer interactions are essential for maintaining network health and connectivity. These interactions enable peers to identify one another, exchange key information, and update routing tables, ensuring efficient communication across the network.

As illustrated in figure~\ref{fig:measure-arch}, the key types of peer interactions include:

\begin{itemize}[leftmargin=*,topsep=1pt]

\item \textbf{Endpoint Proof:} Nodes confirm their availability by responding to \texttt{Ping} messages with \texttt{Pong} replies. This periodic exchange ensures the node is still active and reachable, maintaining the integrity of the distributed hash table (DHT) and preventing traffic amplification attacks.

\item \textbf{Node Discovery Process:} Nodes use \texttt{FindNode} messages to search for peers and expand their network view. The recipient responds with a \texttt{Neighbors} message containing nearby nodes, updating the DHT and facilitating decentralized peer discovery.

\item \textbf{Identity Resolution (ENR):} Nodes exchange \texttt{ENRRequest} and \texttt{ENRResponse} messages to confirm peer identities and share updated public keys and endpoints. This process keeps routing tables accurate and ensures proper protocol compliance.

\item \textbf{Sub-protocol Exchange:} After connecting, nodes exchange a \texttt{Hello} message via the RLPx protocol to share client details. Ethereum nodes then exchange \texttt{Status} messages to confirm compatible network IDs, protocol versions, and fork identifiers. Only peers with matching sub-protocols can communicate further.

\item \textbf{Connection and Disconnection Events:} Nodes manage connections with basic messages like \texttt{Ping} and \texttt{Pong}. If a peer fails to meet protocol conditions or doesn’t respond, the connection may be terminated. These events, governed by peer scoring, ensure the network remains stable and responsive.

\end{itemize}

Analyzing these interactions is key to evaluating the robustness of Ethereum’s discovery and communication protocols, revealing both expected behaviors and potential protocol deviations that may impact network performance.

\section{Example of Peer Information}
The following is an example of peer information extracted from the \texttt{Hello} message:

\begin{lstlisting}[caption=Peer infomation example]
  "id": "ce85a9d87...",
  "ip": "116.203.x.x",
  "udp": 30303,
  "tcp": 30303,
  "clientId": "Geth/v1.13.14-stable-2bd6bd01/linux-amd64/go1.21.7",
  "sub-protocols": ["eth/68", "snap/1"],
  "networkID": (*@{\texttt{7065746802}}@*),
  "genesisHash": "0xec886290bc1f5...",
  "forkID": "ce117a3f/956b78c5c",
\end{lstlisting}

In this example, the peer's information is structured as follows: The \texttt{id} field represents the unique identifier of the peer, derived from its public key. The \texttt{ip}, \texttt{udp}, and \texttt{tcp} fields contain the IP address and port numbers used for communication over their respective protocols. The \texttt{clientId} reveals the client software and version that the peer is running, such as Geth v1.13.14. Furthermore, the \texttt{sub-protocols} field lists the Ethereum sub-protocols that the peer supports, including \texttt{eth/68} for the Ethereum mainnet and \texttt{snap/1} for the snapshot synchronization protocol. The \texttt{networkID} identifies the network to which the peer belongs, while the \texttt{genesisHash} represents the hash of the genesis block that marks the starting point of the blockchain. Finally, the \texttt{forkID} provides a combination of the current chain’s fork hash and the block number for the next scheduled fork, ensuring that the peer is using a compatible protocol version. This data is essential for determining the capabilities and compatibility of peers within the network.

\section{Ethics}

All collected data are anonymized to eliminate any risk of personal identification. Only essential data for research purposes are collected.
Bandwidth usage is limited to 1GBps to minimize the impact on the Ethereum network, and measurement tools are designed not to burden Ethereum notably. In fact, our measurement targets peers and their interactions with us. Only ten messages per minute on average are sent, which is quite acceptable.
These measures ensure adherence to ethical standards, safeguard participant rights and promote network health.
\end{document}